\documentclass[prd,a4paper,nofootinbib]{revtex4}
\usepackage{epsfig}
\usepackage{amsmath}
\usepackage{axodraw}
\usepackage{pstricks}
\usepackage{graphicx}
\usepackage{inputenc}
\usepackage{pst-coil}

\begin{document}
\title{Embedding $A_4$ into $SU(3)\times U(1)$ flavor symmetry:\\
Large neutrino mixing and fermion mass hierarchy in $SO(10)$ GUT}

\author{F.~Bazzocchi$\,^1$, S.~Morisi$\,^1$, M.~Picariello$\,^2$, E.~Torrente-Lujan$\,^3$}
\affiliation{
$^1$ Instituto de F\'{\i}sica Corpuscular -- C.S.I.C./Universitat de Val{\`e}ncia %\\
%  Campus de Paterna, Apt 22085,
E--46071 Val{\`e}ncia, Spain
}

\affiliation{
$^2$    Dipartimento di Fisica - Universit\`a del Salento {\em and} INFN\\%Istituto Nazionale di Fisica Nucleare\\
  Via Arnesano, ex collegio Fiorini, I--73100 Lecce, Italy}

\affiliation{
$^3$	University of Murcia - 30100 Murcia - Spain
\\
\mbox{\\
\em e-mails: bazzocchi@ific.uv.es, morisi@ific.uv.es, Marco.Picariello@le.infn.it, torrente@cern.ch}
}

\begin{abstract}
We present a common explanation of the fermion
mass hierarchy and the large lepton mixing angles in the context of a
grand unified flavor and gauge theory (GUTF).
Our starting point is a $(SU(3) \times U(1))^F$ flavor symmetry and a
$SO(10)$ GUT, a basic ingredient of our theory which plays a major role
is that two different breaking pattern of the flavor symmetry are at work.
On one side, the dynamical breaking of $(SU(3) \times U(1))^F$ flavor
symmetry into $(U(2)\times Z_3)^F$
explains why one family is much heavier than the others.
On the other side, an explicit symmetry breaking of $SU(3)^F$ into
a discrete flavor symmetry  leads to the
observed tribimaximal mixing for the leptons. We write an explicit model where this discrete 
symmetry group is $A_4$.
Naturalness of the charged fermion mass hierarchy appears as a consequence of the continuous 
$SU(3)^F$ symmetry.
Moreover, the same discrete $A_4$-GUT invariant operators are the root of the
large lepton mixing, small Cabibbo angle, and neutrino masses.

\end{abstract}

\maketitle

\section{introduction}

%{\bf TO BE IMPROVED}:

Grand Unified Theory (GUT) \cite{Pati:1973uk,Georgi:1974sy} are natural
extensions of the Standard Model (SM)
Indications toward GUT are the tendency to unify for the gauge couplings,
and the possibility to explain charge quantization and anomaly cancellation.
One of the main features of GUT is its potentiality to unify
the particle representations and the fundamental parameters
in a hopefully predictive framework. 
$SO(10)$ is the smallest simple Lie group for which a
single anomaly-free irreducible representation (namely the spinor
16 representation) can accommodate the entire SM fermion content
of each generation.

Flavor physics appears as  new extra horizontal symmetries.
After the recent experimental evidences about neutrino physics
\cite{SNO,Krauss:2006qq%,Abe:2006fu
,SKatm,%
%SKsolar,
GNO,GALLEX,HOMESTAKE,SAGE,KamLAND,CHOOZ,PaloVerde,Collaboration:2007xf,Michael:2006rx},
within the experimental errors, the neutrino mixing matrix is compatible with the 
so called tri-bimaximal matrix \cite{Harrison:2002er}
%%%%%%%%%%%%%%%%%%%%%%%%%%%%%%%%%%%%%%%%%%%%%%%%%%%
\begin{equation}
U_{TB}=
\left(\begin{array}{ccc}
-2/\sqrt{6} & 1/\sqrt{3} & 0\\
1/\sqrt{6} & 1/\sqrt{3} & 1/\sqrt{2}\\
1/\sqrt{6} & 1/\sqrt{3} & -1/\sqrt{2}\\
\end{array}\right)\,.
\end{equation}
%%%%%%%%%%%%%%%%%%%%%%%%%%%%%%%%%%%%%%%%%%%%%%%%%%%%
At this stage the parameters both the quark \cite{Charles:2004jd}
and lepton \cite{Fogli:2005gs,Aliani:2003ns,Aliani:2002na,Balantekin:2004hi,Oberauer:2004ji,%Rodejohann:2006ek,%
%Gonzalez-Garcia:2006wm,
Valle:2006vb,Fogli:2006jk,Bandyopadhyay:2006jn,Bilenky:2006sn,%
%Messier:2006yg,
Strumia:2006db,%Schwetz:2006dh,Fukugita:2006rm,Chen:2006rk,Robertson:2006pk,
Petcov:2006yg,McDonald:2006qf}
sectors are known to a comparable level.

To explain at the same moment the charged fermion mass hierarchy 
and the lepton-quark mixing angle hierarchy is an unsolved problem,
this is the flavor puzzle.
The problem of the mass hierarchy is often addressed by introducing
continuous flavor symmetries \cite{Barbieri:1999km,Altarelli:2006ri}.
On the other hand, discrete flavor symmetry
 such as 2-3 \cite{23,Balaji:2001ex,Mohapatra:2004mf}, 
$S3$ \cite{Morisi:2005fy,S3,Picariello:2006sp,Caravaglios:2005gw},
$A_4$ \cite{Morisi:2007ft,Bazzocchi:2007au,A4,King:2006np},
or other symmetries \cite{S4,Altarelli:2006kg,Kobayashi:2006wq,Koide:2007kw,Altarelli:2005yp,%
Altarelli:2005yx,Ma:2006vq,Babu:2002dz,Hirsch:2003dr,Grimus:2004rj,Das:2000uk,0802.0090}, %,Hagedorn:2006ug
%, $D_3$, $D_4$ \cite{D3D4}, etc,
where introduced to explain large lepton mixing angles,
but in that case mass hierarchy remains unexplained.

A milestone in these studies has been the discovery that mass
hierarchies and mixing angles can be not directly correlated among them
in the flavor symmetry breaking \cite{Picariello:2006sp,Picariello:2007yn}.
Fundamental steps in the realization of these ideas are
given in \cite{Morisi:2007ft,Bazzocchi:2007au}.
These new ingredients allow us to escape from the no-go theorem \cite{Feruglio:2004gu}
that seems to indicate that a maximal mixing angle $\theta_{23}$
can never arise in the symmetric limit of whatever flavor symmetry
(global or local, continuous or discrete), provided that such a symmetry 
also explains the hierarchy among the fermion masses and is only broken 
by small effects, as we expect for a meaningful symmetry.

In fact, in our theory, the mass hierarchy and large mixing angle
are not originated at the same step in the symmetry breaking pattern.

Our final aim would be the construction of a grand unified $SO(10)$-like
model where masses and mixing angles are generated by the flavor and gauge
symmetry  breaking.

We presented a viable $SO(10)$ model with discrete flavor symmetry
in \cite{Morisi:2007ft}.
There we generated the observed lepton mixing but we fitted the fermion masses
by assuming  the group  $A_4$ as flavor symmetry and the ``constrain'' of
assigning right and left-handed fermion fields to the same representations.
Indeed, we showed in \cite{Morisi:2007ft} that the assignment
of both left-handed and right-handed SM fields to triplets of $A_4$, that is 
therefore compatible with  $SO(10)$,  can lead to  the charged fermion textures
proposed in \cite{Ma:2006wm} and  given by 
%%%%%%%%%%%%%%%%%%%%%%%%%%%%%%%%%%%%%
\begin{equation}\label{eq:A4}
M_f=\left(
\begin{array}{ccc}
h^f_0 & h^f_1& h^f_2\\
h^f_2 & h^f_0 & h^f_1 \\
h^f_1 & h^f_2&h^f_0
\end{array}
\right)\,,
\end{equation}
%%%%%%%%%%%%%%%%%%%%%%%%%%%%%%%%%%%%%%%%%%%
with $h^f_0,\,h^f_1$ and $h^f_2$  distinct parameters. 
In \cite{Morisi:2007ft},
in order to obtain a mass matrix of the form of $M_f$ in eq.~(\ref{eq:A4}) without
spoiling the predictions of the neutrino sector, we introduced higher order operators
containing simultaneously a set of $SO(10)$ representations ${\bf45}$.
The lepton mixing was naturally generated by the breaking pattern of $A_4$, while the
fermion masses were obtained with a possible tuning in the flavor parameters not
constrained by the symmetries. 

We addressed the problem of the fine tuning in \cite{Bazzocchi:2007au}
where the $A_4^F$ flavor discrete symmetry is embedded into 
$(SO(3)_L\times SO(3)_R)^F$. In that way  we explicitly disentangled the mixing problem
from the hierarchy one.
We broke the continuous flavor $(SO(3)_L\times SO(3)_R)^F$ symmetry both dynamically and
explicitly. The two breaking terms produced the charged fermion hierarchies
on one hand and solved the leptonic mixing problem on the other hand. 
In this way not only a tribimaximal neutrino mixing was naturally generated but
also the charged fermion hierarchies by dynamically breaking of the continuous
left-right flavor symmetry. Finally the Cabibbo angle was obtained by taking
into account higher order operators.
However the left-right flavor group symmetry $(SO(3)_L\times SO(3)_R)^F$ of
\cite{Bazzocchi:2007au} is not compatible with a grand unified gauge group, like $SO(10)$,
with all the fermions of one family in the same representation, because in left-right
flavor symmetries the fermions of one family belong to different representations
of the flavor group.
 
In this paper we merge all these ingredients together and we are able
to construct a non renormalizable model with grand unified gauge group
$SO(10)$ and with an extended flavor symmetry $(SU(3) \times U(1))^F$. In this new
model both the tribimaximal lepton mixing matrix and the hierarchy
among the mass of the 3rd and the other fermion families naturally appear
from the symmetry breaking pattern.
Our model is non renormalizable, however a renormalizable version of it can be
easily constructed because the particular structure  of the operators  introduced  here.
For this purpose viable methods are well known, i.e. by integrated out given heavy extra fields 
\cite{%Malinsky:2007qy,McKeen:2007ry,
Berezhiani:1996bv}.%,Barr:2007ma}.

Our effective $SO(10)$ invariant Lagrangian is
%%%%%%%%%%%%%%%%%%%%%%%%%%%%%%%%%%%%%
\begin{eqnarray}\label{eq:L1}
\mathcal{L}&=&\mathcal{L}_{SU(3)^F}+\delta \mathcal{L}_{A_4}\,,
\end{eqnarray}
%%%%%%%%%%%%%%%%%%%%%%%%%%%%%%%%%%%%%
where $\mathcal{L}_{SU(3)^F}$ is $SO(10)\times (SU(3) \times U(1))^F$ invariant and 
$\delta \mathcal{L}_{A_4}$ is the explicit breaking term of the $SU(3)^F$ 
symmetry that, at this level, leaves $SO(10)$ unbroken.
The charge assignment of the fields is such that the $SU(3)^F$ invariant
operator with lowest mass dimensions is only \cite{Morisi:2007ft}
%%%%%%%%%%%%%%%%%%%%%%%%%%%%%%%%%%%%%
\begin{eqnarray}\label{eq:LagU3}
\mathcal{L}_{SU(3)^F}&=&h_0 \,{\bf 16}\,{\bf 10}\,{\bf 45}_{A}\,{\bf 45}_B\,{\bf 16}\, {\Phi}\,,
\end{eqnarray}
%%%%%%%%%%%%%%%%%%%%%%%%%%%%%%%%%%%%%
where ${\Phi}$, singlet of $SO(10)$, transforms as $\overline{\bf 6}$
with respect to $SU(3)^F$. The scalar fields
${\bf 10}$, ${\bf 45}_A$ and ${\bf 45}_B$ are singlets of $SU(3)^F$.
As noticed in \cite{Morisi:2007ft}, thanks to the two ${\bf 45}$s scalar fields,
the operator in eq. (\ref{eq:LagU3}) can give no contribution to the neutrino 
sector for some set of the ${\bf 45}$s vev (i.e. in the explicit model of \cite{Morisi:2007ft}
one vev is proportional to the right-handed isospin $T_{3_R}$ and the other one to the
hypercharge $Y$, here the directions $A$ and $B$ are different but still with the same property).

When the scalar field $\Phi$ develops a vev  in the direction
$\langle\Phi^{ij}\rangle=1\, (\forall\, i\,,j)$  we obtain a democratic
mass matrices for all the charged fermions, that gives a massive 
3rd family and two massless families. The democratic structure of the 
charged fermion mass matrices avoids the fine tuning needed to explain the mass
hierarchy between the 3rd and the other two families that is usually 
needed in presence of charged fermion mass matrices of the form of eq. (\ref{eq:A4}). 
The democratic mass matrices preserves the $(U(2)\times Z_3)^F$ subgroup of
$(SU(3) \times U(1))^F$ that leaves invariant
the $(1,1,1)$ vector in the flavor space.
Therefore at this stage only one mixing angle can be generated.

The neutrino mass matrix and the first and second families masses arise
when we switch on the  explicitly breaking terms of $SU(3)^F$ into $A_4$.
If we neglect the ordering problem of the {\bf 45}s and the possibility to have more than one flavon for each
operator, the most general Lagrangian invariant under the flavor structure of the theory is
%%%%%%%%%%%%%%%%%%%%%%%%%%%%%%%%%%%%%
\begin{eqnarray}\label{eq:LagA4}
\delta \mathcal{L}_{A_4}&=&
  h_{ijk}  ~\phi^k~{\bf16}^i~{\bf 10}~{\bf45}_A~{\bf45}_B~{\bf45}_{C}~{\bf45}_D~{\bf16}^j+
  h_{ijk}' ~\phi^k~{\bf16}^i~{\bf45}_{C}~{\bf45}_D~{\bf 10}~{\bf45}_A~{\bf45}_B~{\bf16}^j+\\\nonumber&&
  h_{ijk}''~\tilde{\phi}^k~{\bf16}^i~{\bf 10}~{\bf45}_{C}~{\bf45}_D~{\bf16}^j
 +  g~       {\bf16}^i~\overline{\bf126}%~{\bf45}_A~{\bf45}_B
 ~{\bf16}^i\,\zeta_S
 +  g'_{ijk}~{\bf16}^i~\overline{\bf126}%~{\bf45}_A~{\bf45}_B
 ~{\bf16}^j\,\zeta_T^k
\end{eqnarray}
%%%%%%%%%%%%%%%%%%%%%%%%%%%%%%%%%%%%%
where the indices $\{i,j,k,l\}$ are  $A_4$, subgroup of $SU(3)$, indeces
 and the sum over the gauge indices is understood.
The scalar field $\overline{\bf 126}$ is a singlet
${\bf 1'}$ of $A_4$, while the ${\bf 45}_C$, and ${\bf 45}_D$
are other scalars that transform as {\bf 45} of $SO(10)$, 
and are singlets of $A_4$.
The flavon fields $\phi$, $\tilde\phi$, $\zeta_T$ are triplets under $A_4$, while
$\varphi$ and $\zeta_S$ are singlets.

As found in \cite{Morisi:2007ft} the terms in the second line of 
$\delta \mathcal{L}_{A_4}$ generates the light neutrino mass matrix.
The terms in the first two lines in $\delta \mathcal{L}_{A_4}$
gives a contribution to the mass matrices
that has the nice properties to commute with the leading order
term obtained from eq.~(\ref{eq:L1}).

After the breaking of $A_4$, it generates the first and second
family masses and fix the mixing matrix in the lepton sector to
be tribimaximal.

The plan of the paper is as follow.
First, in sec. {\bf\ref{sec:basic}} we introduce the basic ingredient
of the model, i.e. the general structure of the symmetry breaking, 
all the involved fields and how they transform under the gauge
and flavor symmetries.
Then in sec. {\bf\ref{sec:breaking}} we show how the 3rd family masses
are generated via the breaking $(SU(3)\times U(1))^F$ into $(U(2)\times Z_3)^F$, how
the 1st and 2nd family masses are generated together with maximal
mixings in the lepton sector, and how the neutrino masses are
generated with a resulting tribimaximal mixing matrix in the
lepton sector. Finally we show how the Cabibbo angle is naturally
generated without the introduction of new operators.
Finally in sec. {\bf\ref{sec:conclusions}} we report our conclusions.
%We also added an appendix {\bf\ref{sec:appendice}} to implement
%the idea of \cite{Anderson:1993fe} to transform our effective
%theory into a renormalizable one, by including a set
%of heavy spinor fields and extra symmetries.

\section{Basic ingredients}\label{sec:basic}
Let us first investigate the field content of the theory and the flavor
charges.
We report the field content of our model in Table (\ref{tab:tab1}).
With our charge assignment, the only allowed operators of lower mass dimensions 
are given in eqs. (\ref{eq:LagU3}-\ref{eq:LagA4}), if we neglect the ordering problem of the $\bf 45$
and the possibility to have more than one flavon for each operator.
In this sense our Lagrangian is the most general one invariant under the flavor structure of
the theory.
Moreover, independently from the fact that nature prefer a dominant seesaw of
type I (i.e. heavy Majorana right-handed neutrino mass and intermediate Dirac neutrino mass)
or of type II (i.e. light Majorana left-handed neutrino mass) or a mixed scenario,
the transformation properties of the $\zeta_S$ must be assumed to be ${\bf 1}'$,
as we will explain in sec. {\bf\ref{sec:neutrino}}. 

In our opinion, the ordering problem can be related to a deeper structure of the theory, for example
its version as a renormalizable model, and we will not investigate further it here. However the fact
that will not be possible to express the directions $A$ and $B$ as rational combinations of $C$ and $D$,
together with the fact that ${\bf 45}$ appears only as couples (${\bf 45}_A$,${\bf 45}_B$) and
(${\bf 45}_C$,${\bf 45}_D$) seems to us to indicate that the right representations to introduce are
the irreducible part of the ${\bf 2025}$ that can get a vev diagonal over the ${\bf 16}$ matter fields
with charges $A B$ and $C D$.
If this is the case we are really including all the allowed operator and there is not any more
an ordering problem.

After symmetry breaking, 
once the Higgs acquire vevs, the quadratic part for the fermions
of the Lagrangian in eqs. (\ref{eq:LagU3}-\ref{eq:LagA4}) can be rewritten
in a compact form, i.e. with an abuse of notation in the $SO(10)$ contractions,
as
\begin{subequations}\label{eq:quadratic}
\begin{eqnarray}
L_\text{Dirac}&=&
h_0~({\bf16}_1{\bf16}_1'+{\bf16}_2{\bf16}_2'+{\bf16}_3{\bf16}_3')~v_{\bf 10}+
\\&+&
\Big[h_1~({\bf16}_1{\bf16}_2'''+{\bf16}_2{\bf16}_3'''+{\bf16}_3{\bf16}_1''')
+h_2~({\bf16}_1{\bf16}_3'''+{\bf16}_2{\bf16}_1'''+{\bf16}_3{\bf16}_2''')\Big]~v_{\bf 10}~v_{\bf \phi}
\\&+&
\Big[h'_1~({\bf16}_1''{\bf16}_2'+{\bf16}_2''{\bf16}_3'+{\bf16}_3''{\bf16}_1')%~v_{\bf 10}~v_{\bf \phi}
+h'_2~({\bf16}_1''{\bf16}_3'+{\bf16}_2''{\bf16}_1'+{\bf16}_3''{\bf16}_2')\Big]~v_{\bf 10}~v_{\bf \phi}\\
&+&
\Big[h''_1~({\bf16}_1{\bf16}_2''+{\bf16}_2{\bf16}_3''+{\bf16}_3{\bf16}_1'')%~v_{\bf 10}~v_{\bf \tilde\phi}
+h''_2~({\bf16}_1{\bf16}_3''+{\bf16}_2{\bf16}_1''+{\bf16}_3{\bf16}_2'')\Big]~v_{\bf 10}~v_{\bf \tilde\phi}\\
%L_\text{Majo}&=&
%&+& \lambda~{\bf16}^i~{\bf16}^i~v_{\bf10}\,v_\varphi\\\nonumber
&+&g({\bf16}_1{\bf16}_1+{\bf16}_2{\bf16}_2+{\bf16}_3{\bf16}_3)~v_{\bf\overline{126}}~v_{\zeta_S}
+\Big[g'_1{\bf16}_1{\bf16}_2%~v_{\bf126}~v_{\zeta_T}
+g'_2{\bf16}_2{\bf16}_1\Big]~v_{\bf\overline{126}}~v_{\zeta_T}
\end{eqnarray}
\end{subequations}
where we have assumed that the two $A_4$-${\bf3}$plets $\phi$ and $\tilde\phi$ acquire vev in the 
$(1,1,1)$ direction
of $A_4$, while the $\zeta_T$ vev is in the direction $(0,0,1)$.
In eqs. (\ref{eq:quadratic}) we introduced
\begin{eqnarray}
%{\bf16}_i'''\equiv v_{{\bf45}_{A}}~v_{{\bf45}_B}~v_{{\bf45}_{T_{3R}}}~v_{{\bf45}_Y}~{\bf16}_i
%&\quad\quad&
%{\bf16}_i''\equiv v_{{\bf45}_{T_{3R}}}~v_{{\bf45}_Y}~{\bf16}_i\\
{\bf16}_i'''\equiv v_{{\bf45}_{A}}~v_{{\bf45}_B}~v_{{\bf45}_C}~v_{{\bf45}_D}~{\bf16}_i\,,
&\quad\quad&
{\bf16}_i''\equiv v_{{\bf45}_C}~v_{{\bf45}_D}~{\bf16}_i\,,\\
{\bf16}_i'\equiv v_{{\bf45}_{A}}~v_{{\bf45}_B}~{\bf16}_i\,,
&\quad\quad&
\text{with}~~i=1,2,3\,.\nonumber
\end{eqnarray}
 We obtain the following expression by absorbing 
  the vevs of the ${\bf 45}$s into the coupling constants %$h_0$, $h$, $h'$, $h''$
\begin{subequations}
\begin{eqnarray}
{\bf 16'}&=&
\left(
{{x'_Q}}~Q,
~{{x'_U}}~U^c,
~{{x'_D}}~D^c,
~{{x'_L}}~L,
~{{x'_E}}~E^c,
~{{x'_N}}~N^c
\right),
\\
{\bf 16''}&=&
\left(
{{x''_Q}}~Q,
~{{x''_U}}~U^c,
~{{x''_D}}~D^c,
~{{x''_L}}~L,
~{{x''_E}}~E^c,
~{{x''_N}}~N^c
\right),
\\
{\bf 16}'''&=&
\left(
{{x'''_Q}}~Q,
~{{x'''_U}}~U^c,
~{{x'''_D}}~D^c,
~{{x'''_L}}~L,
~{{x'''_E}}~E^c,
~{{x'''_N}}~N^c
\right)
\end{eqnarray}
\end{subequations}
where $x'_f$, $x''_f$, and $x'''_f$ are the
quantum numbers respectively of the product of the charges $A$ and $B$, of
the product of the charge %$T_{3R}$ with $Y$,
$C$ and $D$, and of the product
of the charges %$A$, $B$, ${T_{3R}}$, and $Y$.
$A$, $B$, $C$, and $D$.
In particular we notice that
\begin{equation}\label{eq:times}
x'''_f=x''_f x'_f\,.
\end{equation}
We report the charges of each fermion in Table (\ref{tab:tab2}).

\section{Dynamical breaking}\label{sec:breaking}

\subsection{ $(SU(3)\times U(1))^F\rightarrow (U(2)\times Z_3)^F$ gives the charged fermion 3rd family masses}\label{sec:dynamical}

We assume that the $\Phi$ $SU(3)^F$-$\overline{\bf 6}$plet field acquire a vev % in the direction
$\langle\Phi^{ij}\rangle= v_\Phi \,(\forall\,i\,,j)$.
In this case the charged fermion mass matrix obtained is the so-called
democratic mass matrix \cite{Harari:1978yi} given by
\begin{equation}\label{eq:massdem}
M_{0f}=\frac{m_3^f}{3}\left(
\begin{array}{ccc}
1&1&1\\
1&1&1\\
1&1&1
\end{array}
\right)\,,
\end{equation}
where
\begin{subequations}
\begin{eqnarray}
m_3^U&=&v_\Phi (x'_{U} + x'_{Q}) v_{\bf 10}^U h_0\,,\\
m_3^D&=&v_\Phi (x'_{D} + x'_{Q}) v_{\bf 10}^D h_0\,,\\
m_3^N&=&v_\Phi (x'_{N} + x'_{L}) v_{\bf 10}^U h_0\,,\\
m_3^E&=&v_\Phi (x'_{E} + x'_{L}) v_{\bf 10}^D h_0\,.
\end{eqnarray}
\end{subequations}
This matrix has only one eigenvalue different from zero, $m_3^f$, and can 
be assumed to be the mass  of the 3rd family. 
%Let's call $Q_A(\psi)$ and $Q_B(\psi)$ the charge of the ${\bf 45}$s
%vev on the field $\psi$.
%can be written as
%$Q_{A,B}=\alpha_{A,B} Y + \beta_{A,B} T_{3_R}$.
To avoid any non diagonal contribution to the Dirac neutrino mass matrix
we impose 
\begin{subequations}\label{eq:U3cond}
\begin{eqnarray}
x'_N+x'_L&=&0\,.
\end{eqnarray}
 %$Q_A(\nu) Q_B(\nu) + Q_A(L) Q_B(L)=0$.
%, and the other direction undetermined at the moment.
%With respect to \cite{Morisi:2007ft}, the fact that only one contraction
%is allowed in the flavor space allows us to have zero contribution to the
%Dirac neutrino mass by imposing one condition only.
%However, t
To have the bottom-tau unification, we must impose also
%%%%%%%%%%%%%%%%%%%%%%%%%%%%%%%%%%%%%%%%%%%%%%%%%%%%%%%%%%%%%%%%%%%%
\begin{eqnarray}
x'_L+x'_E&=&x'_Q+x'_D\,.	
\end{eqnarray}
\end{subequations}
%%%%%%%%%%%%%%%%%%%%%%%%%%%%%%%%%%%%%%%%%%%%%%%%%%%%%%%%%%%%%%%%%%%%%
The unitary matrix $U$ that diagonalizes the symmetric matrix $M_{0f}$ has 
one angle and the three phases undeterminated. One possible 
parametrization is given by \cite{Bazzocchi:2007au}
\begin{equation}
\label{eq:Uparam}
U=
\frac{1}{\sqrt{3}}\,\left(
\begin{array}{ccc}
 \sqrt{2}\cos\theta\,e^{i \alpha}
&\sqrt{2}\sin\theta e^{i (\beta+\gamma)}
&1\\
-e^{i\alpha}(\frac{\cos\theta}{\sqrt{2}}+
             \sqrt{\frac{3}{2}} \sin\theta e^{-i\gamma})
&e^{i \beta}(\sqrt{\frac{3}{2}}\cos\theta
             -\frac{1}{\sqrt{2}}\sin\theta e^{i \gamma})
&1\\
-e^{i \alpha}(\frac{\cos\theta}{\sqrt{2}}
             -\sqrt{\frac{3}{2}}\sin\theta e^{-i\gamma})
&-e^{i \beta}(\sqrt{\frac{3}{2}}\cos\theta
              +\frac{1}{\sqrt{2}}\sin\theta e^{i \gamma})
&1
\end{array}
\right)\,.
\end{equation}
The freedom in the $U$ matrix shows the remaining flavor symmetry $U(2)^F$.
The unknow angle and phases are fixed only after breaking the 
democratic structure of $M_{0f}$, i.e. the $U(2)^F$ flavor symmetry,
with a small perturbation $\delta M_f$, i.e. 
$$ M_f=M_{0f}+\delta M_f\,.$$
The effect of $\delta M_f$ is to give a small mass to the first and 
second family and to fix the mixing angles.
To have that the mixing matrix diagonalizing the full $M_f$ belongs
to the families of matrix of eq. (\ref{eq:Uparam}), we must require
that $\delta M_f$ commute with $M_{0f}$. This has the nice consequence
that we have automatically the selection of the breaking pattern
of $A_4$ into $Z_3$.

\section{Explicitly breaking $SU(3)^F \rightarrow A_4$}
We will assume the presence of an hidden scalar sector that 
breaks spontaneously  the continuous $SU(3)^F$ into the discrete $A_4$.
Under this hypothesis it is quite  natural to assume that the explicit breaking
terms %in eq.~(\ref{eq:L1}), 
to be added to the Lagrangian %of eq.~(\ref{eq:L0}),
are small.

\subsection{$A_4\to Z_3$ generates the charged fermions 1st and 2nd family masses and mixing}
When the $\phi$ and $\tilde\phi$ $A_4$-${\bf 3}$plets take vev as 
$\langle \tilde\phi \rangle\propto\langle \phi \rangle=v_\phi\,(1,1,1)$
we have new contributions to the mass matrices. I.e., for the charged leptons
we get the operator
\begin{equation}\label{eq:epsi}
\delta^E_{ijk}\,\epsilon_{\alpha\beta}\,H_d^\alpha 
\,\left(L_i^\beta\,E_j\,\phi_k\right)
\rightarrow \epsilon_{\alpha\beta}\,H_d^\alpha
\left[\delta^E_1 \,(L_2^\beta E_3+L_3^\beta E_1+L_1^\beta E_2)+ 
\delta^E_2 \,(L_3^\beta E_2+L_1^\beta E_3+L_2^\beta E_1)\right] v_\phi\,,
\end{equation}
where the two $\delta^E_i$ arise by the two different contractions of $A_4$.
The value of $\delta^E$s can be read from the Lagrangian in eq.
(\ref{eq:quadratic}) and is
%%%%%%%%%%%%%%%%%%%%%%%%%%%%%%%%%%%%%%%%%%%%%%%%%%%%%%%%%
\begin{eqnarray}\label{eq:deltaE}
\delta_1^E &=& 
(h_1 x_E''' + h_2 x_L''')+
(h_1' x_L''x_E' + h_2' x_L'x_E'')+
(h_1'' x_E'' + h_2'' x_L'') \,,
\\
\delta_2^E &=& 
(h_2 x_E''' + h_1 x_L''')+
(h_2' x_L''x_E' + h_1' x_L'x_E'')+
(h_2'' x_E'' + h_1'' x_L'') \,.
\end{eqnarray}
%%%%%%%%%%%%%%%%%%%%%%%%%%%%%%%%%%%%%%%%%%%%%%%%%%%%%%%%%
Because the $SO(10)$ unification, the operators in the up, down
and neutrino sectors have similar expressions.
In particular for the contributions to the  Dirac neutrino mass
matrix we have
%%%%%%%%%%%%%%%%%%%%%%%%%%%%%%%%%%%%%%%%%%%%%%%%%%%%
\begin{eqnarray}\label{eq:deltaN}
\delta_1^N &=& 
(h_1 x_N''' + h_2 x_L''')+
(h_1' x_L''x_N' + h_2' x_L'x_N'')+
(h_1'' x_N'' + h_2'' x_L'') \,,
\\
\delta_2^N &=& 
(h_2 x_N''' + h_1 x_L''')+
(h_2' x_L''x_N' + h_1' x_L'x_N'')+
(h_2'' x_N'' + h_1'' x_L'') \,.
\end{eqnarray}
%%%%%%%%%%%%%%%%%%%%%%%%%%%%%%%%%%%%%%%%%%%%%%%%%%%%%%%%%
At this stage we do not want any contribution to the Dirac neutrino mass
matrix, without any fine tuning in the coupling constants $h$s.
For this reason we have to impose the conditions
%\begin{subequations}
\begin{eqnarray}\label{eq:A4cond}
x'''_L=0=x'''_N\,,\quad
x''_L x'_N=0=x'_L x''_N\,,\quad
x''_L=0=x''_N \,.
\end{eqnarray}
%\end{subequations}
%We also remember here the conditions coming
%from the $m_\tau$-$m_b$ unification of sec. {\bf\ref{sec:dynamical}}
%\begin{eqnarray}
%x'_N+x'_L=0\,,\quad
%x'_E+x'_L=x'_D+x'_Q\,.
%\end{eqnarray}
%\end{subequations}
By reducing the set of conditions in eqs. (\ref{eq:times}),
 (\ref{eq:U3cond}),
 and (\ref{eq:A4cond})
the only non trivial solution is
\begin{subequations}
\begin{eqnarray}
x'&=&	A\,B \propto 3\, (Y^2) - 12\, (T_{3_R}^2) + 20\, T_{3_R} Y
\\
x''&=&	C\,D \propto (Y\,T_{3_R})
\\
x'''&=&A\,B\,C\,D=x'' x'
\end{eqnarray}
\end{subequations}
%%%%%%%%%%%%%%%%%%%%%%%%%%%%%%%%%%%%%%%%%%%%%%%%%%%%%%%%%%%%%
In particular we notice that we have
%%%%%%%%%%%%%%%%%%%%%%%%%%%%%%%%%%%%%%%%%%%%%%%%%%%%%%%%%%%%%
\begin{equation}\label{eq:zeros}
x''_Q=x'''_Q=x''_L=x'''_L=x''_N=x'''_N=0\,.
\end{equation}
%%%%%%%%%%%%%%%%%%%%%%%%%%%%%%%%%%%%%%%%%%%%%%%%%%%%%%%%%%%%%
Finally, by taking into account the relations in eq. (\ref{eq:zeros}),
and the relation (\ref{eq:times}), we get from eq. (\ref{eq:deltaE})
%%%%%%%%%%%%%%%%%%%%%%%%%%%%%%%%%%%%%%%%%%%%%%%%%%%%%%%%%%%%%
\begin{subequations}
\begin{eqnarray}
\delta_1^E = x''_E (x'_E h_1+x'_L h_2' + h_1'')\,,&\quad&
\delta_2^E = x''_E (x'_E h_2+x'_L h_1' + h_2'')\,,\\
\delta_1^U = x''_U (x'_U h_1+x'_Q h_2' + h_1'')\,,&\quad&
\delta_2^U = x''_U (x'_U h_2+x'_Q h_1' + h_2'')\,,\\
\delta_1^D = x''_D (x'_D h_1+x'_Q h_2' + h_1'')\,,&\quad&
\delta_2^D = x''_D (x'_D h_2+x'_Q h_1' + h_2'')\,,\\
\delta_1^N = 0\,,&\quad&
\delta_2^N = 0\,.
\end{eqnarray}
\end{subequations}
%%%%%%%%%%%%%%%%%%%%%%%%%%%%%%%%%%%%%%%%%%%%%%%%%%%%%%%%%%%%%%%%%%
We notice here the importance of having the three operators.
In fact, for example, if we had only 
%$  h_{ijk}~\phi^k~{\bf16}^i~{\bf 10}~{\bf45}_{T_{3R}}~{\bf45}_Y~
%  {\bf45}_C~{\bf45}_D~{\bf16}^j$
%, who produce at effective level the operator in (\ref{eq:epsi}), 
one than we should obtain the relations
\begin{eqnarray}
\frac{m_e}{m_\mu} = \frac{m_d}{m_s} = \frac{m_u}{m_c}\,.
\end{eqnarray}
While the first relation can be assumed true at the unification scale, with
the given uncertainty in the determination of the fermion masses at such scale,
the second relation is surely false.
The introduction of the other operators 
allows us to escape from this consequence.
We notice that there could be a direct relation between the fact that
$\frac{m_d}{m_s} \neq \frac{m_u}{m_c}$ and the presence of a non
zero CP-violating phase. 
The effect of the other explicit breaking terms in the mass matrices 
is translated in a perturbation of the democratic mass matrices
of eq.~(\ref{eq:massdem}), that is 
%%%%%%%%%%%%%%%%%%%%%%%%%%%%%%%%%%%%%%%%%%%%%%%%%%%%%%%%
\begin{equation}\label{break1}
 M^f  =\frac{m_3^f}{3}\left(
\begin{array}{ccc}
1&1&1\\
1&1&1\\
1&1&1
\end{array}
\right)
\to
\tilde{M}^f=\,\frac{m_3^f}{3}\left(
\begin{array}{ccc}
1           &1+\delta^f_1&1+\delta^f_2\\
1+\delta^f_2&1           &1+\delta^f_1\\
1+\delta^f_1&1+\delta^f_2&1           
\end{array}
\right)
\end{equation}
with the obvious correspondences $v_E=v_D$.
The mass matrices of eq.~(\ref{break1}) are diagonalized by
%%%%%%%%%%%%%%%%%%%%%%%%%%%%%%%%%%%%%%%%%%%%%%%%%%%%%%%%%%%
\begin{equation}\label{eq:U}
\tilde{U}_\omega=
\frac{1}{\sqrt{3}}\left(
\begin{array}{ccc}
\omega&\omega^2&1\\
\omega^2&\omega&1\\
1&1&1
\end{array}
\right)\,,
\end{equation}
%%%%%%%%%%%%%%%%%%%%%%%%%%%%%%%%%%%%%%%%%%%%%%%%%%%%%%%%%%%%%%%%
corresponding to the $U$ of eq.~(\ref{eq:Uparam}) with $\theta=\pi/4$, $\alpha=2 \pi/3$, $\beta=5 \pi/6$ and $\gamma=\pi/2$.
The mass matrices $\tilde{M}^f$ of eq.~(\ref{break1}) give an heavy 3rd family
mass $m_3^f$ and small 1st and 2nd family masses satisfying
\begin{equation}\label{13}
\frac{m^f_1}{m_3^f}=\frac{\omega\,\delta^f_1+\omega^2\,\delta^f_2}{3+\delta^f_1+\delta^f_2}\,,
\qquad\frac{m^f_2}{m_3^f}=\frac{\omega^2\,\delta^f_1+\omega\,\delta^f_2}{3+\delta^f_1+\delta^f_2}\,.
\end{equation}

\subsection{Neutrino masses and mixing}\label{sec:neutrino}% come from $Z_3\to Z_2$}
\label{sec:newneut}
The Yukawa interactions for the neutrinos come from the
coupling of the fermion field $\bf 16$ with the
$\bf \overline{126}$ Higgs and the ($\zeta_S,\zeta_T$) flavons.
The components of the $\bf \overline{126}$ that can acquire a vev%
%%%%%%%%%%%%%%%%%%%%%%%%%%%%%%%%%%%%%%%%%%%%%%%%%%%%%%%%%%%%%%%%%
\footnote{We neglect here any contribution from the 
$({\bf1},{\bf1},{\bf6})$-plet of the Pati-Salam subgroup of $SO(10)$.}
%%%%%%%%%%%%%%%%%%%%%%%%%%%%%%%%%%%%%%%%%%%%%%%%%%%%%%%%%%%%%%%%%%
are a triplet $\Delta$, three singlets%
%%%%%%%%%%%%%%%%%%%%%%%%%%%%%%%%%%%%%%%%%%%%%%%%%%%%%%%%%%%%%%%%%%%
\footnote{The three singlets $\tilde{\Delta}$ and the two singlets 
$\tilde{\Gamma}$ are respectively a triplet and a doublet under the $SU(2)_R$
of the Pati-Salam group.}
%%%%%%%%%%%%%%%%%%%%%%%%%%%%%%%%%%%%%%%%%%%%%%%%%%%%%
$\tilde{\Delta}$, a doublet $\Gamma$ and two other singlets$^{2}$ $\tilde{\Gamma}_\alpha$
of the weak $SU(2)_L$. 
When the $A_4$-triplet field $\zeta_T$ takes vev in the $A_4$ direction
$\langle \zeta_T \rangle\sim(0,0,1)$ - notice that this alignment is 
different from the one used in many models as for example in
\cite{Morisi:2007ft,Altarelli:2005yp} -,
the resulting neutrino mass matrices are given by
%%%%%%%%%%%%%%%%%%%%%%%%%%%%%%%%%%%%%%%%%%%%%%%%%%%%%%%%
\begin{eqnarray}\label{eq:neutrino}
M_{\nu\nu}=\left(
\begin{array}{ccc}
a_{\nu\nu} & b_{\nu\nu}& 0\\
b_{\nu\nu} &\omega\, a_{\nu\nu} & 0\\
0 &0 &\omega^2\, a_{\nu\nu} 
\end{array}
\right)
\,,\quad\quad
M_{\nu N}=\left(
\begin{array}{ccc}
a_{\nu N} & b_{\nu N}& 0\\
b_{\nu N} &\omega\, a_{\nu N} & 0\\
0 &0 &\omega^2\, a_{\nu N}
\end{array}
\right)
\,,\quad\quad
M_{N N}=\left(
\begin{array}{ccc}
a_{N N} & b_{N N}& 0\\
b_{N N} &\omega\, a_{N N} & 0\\
0 &0 &\omega^2\, a_{N N}
\end{array}
\right)
\,,
\end{eqnarray}
where $a$s and $b$s are the product of the vevs of the $\bf\overline{126}$
components with the coupling constants $g$ and $g'$.
%%%%%%%%%%%%%%%%%%%%%%%%%%%%%%%%%%%%%%%%%%%%%%%%%%%%%%%%%%%%%%%%%5
All the mass matrices in eq. (\ref{eq:neutrino}) are diagonalized
by the same mixing matrix. I.e. we get 
%%%%%%%%%%%%%%%%%%%%%%%%%%%%%%%%%%%%%%%%%%%%%%%%%%%%%%%%%%%
\begin{eqnarray}
M_{x}=\left(
 \begin{array}{ccc}
  a_{x} & b_{x}& 0\\
  b_{x} &\omega\, a_{x} & 0\\
  0 &0 &\omega^2\, a_{x} 
 \end{array}
\right)
=
\tilde{V}_\nu^\star
\left(
 \begin{array}{ccc}
  \omega^2\, a_{x} +b_{x}& 0& 0\\
  0 &\omega^2\, a_{x} & 0\\
  0 &0 &-\omega^2\,a_{x}+ b_{X} 
 \end{array}
\right)\tilde{V}_\nu^\dagger
\end{eqnarray}
%%%%%%%%%%%%%%%%%%%%%%%%%%%%%%%%%%%%%%%
with $x\in\{\nu\nu,\nu N,N N\}$. The common mixing matrix
$\tilde V_\nu$ is given by
%%%%%%%%%%%%%%%%%%%%%%%%%%%%%%%%%%%%%%%
\begin{eqnarray}\label{eq:Vtilde}
\tilde{V}_\nu=\left(
 \begin{array}{ccc}
  \frac{\omega}{\sqrt{2}} &0 & \,-i\frac{\omega}{\sqrt{2}} \\
  \frac{\omega^2}{\sqrt{2}} &0 & i\,\frac{\omega^2}{\sqrt{2}}\\
  0&1&0
 \end{array}
\right)\,.
\end{eqnarray}
%%%%%%%%%%%%%%%%%%%%%%%%%%%%%%%%%%%%%%%%%%%%%%%%%%%%%%%%%%%
The fact that all the mass matrices are diagonalized by the same
mixing matrix $\tilde{V}_\nu$, translates in the nice result that, independently
on the seesaw mechanism acting to generate the low energy
neutrino mass, the neutrino mixing matrix is given by $\tilde{V}_\nu$
itself.
In fact we have
%%%%%%%%%%%%%%%%%%%%%%%%%%%%%%%%%%%%%%%%%%%%%%%%%%%%%%%%%%%
\begin{eqnarray}
	M_{low}&=& M_{\nu\nu} + M_{\nu N}\frac{1}{M_{NN}}M_{\nu N}^T
\end{eqnarray}
%%%%%%%%%%%%%%%%%%%%%%%%%%%%%%%%%%%%%%%%%%%%%%%%%%%%%%%%%%%
and consequently, by indicating with an index $^{\Delta}$
the corresponding diagonalized matrix, we get
%%%%%%%%%%%%%%%%%%%%%%%%%%%%%%%%%%%%%%%%%%%%%%%%%%%%%%%%%%%
\begin{eqnarray}
\tilde V_\nu^T M_{low} \tilde V_\nu&=&
      \tilde V_\nu^T M_{\nu\nu} \tilde V_\nu
    + \tilde V_\nu^T M_{\nu N} (\tilde V_\nu \tilde V_\nu^\dagger)
            \frac{1}{M_{NN}} 
                               (\tilde V_\nu^\star \tilde V_\nu^T) 
            M_{\nu N}^T \tilde V_\nu\nonumber\\	
&=& \tilde V_\nu^T M_{\nu\nu} \tilde V_\nu
    + \tilde V_\nu^T M_{\nu N} \tilde V_\nu 
             \frac{1}{\tilde V_\nu^T M_{NN} \tilde V_\nu}
          \tilde V_\nu^T M_{\nu N}^T \tilde V_\nu\nonumber\\
&=& M_{\nu\nu}^{\Delta}+ M_{\nu N}^{\Delta} \frac{1}{M_{NN}^{\Delta}} M_{\nu N}^{\Delta}\nonumber\\
&=& M_{low}^{\Delta}\,,
\end{eqnarray}
%%%%%%%%%%%%%%%%%%%%%%%%%%%%%%%%%%%%%%%%%%%%%%%%%%%%%%%%%%%
where we inserted twice the identity matrix
$(\tilde V_\nu \tilde V_\nu^\dagger)=1=(\tilde V_\nu^\star \tilde V_\nu^T)$.
The result is that the neutrino mass matrix is diagonalized
by the mixing matrix $\tilde V_\nu$.
On the other hand, the charged leptons are diagonalized by $L\to \tilde{U}_\omega\,L$, so we obtain
a tribimaximal mixing for the lepton sector, that is 
%%%%%%%%%%%%%%%%%%%%%%%%%%%%%%%%%%%%%%%%%%%%%%%%%%%%%%%%%%%
\begin{equation}\label{eq:Vleptons}
V_{leptons}=\tilde{U}^\dagger\cdot \tilde{V}_\nu=
\left(
\begin{array}{ccc}
\frac{2}{\sqrt{6}} &\frac{1}{\sqrt{3}} &0 \\
-\frac{1}{\sqrt{6}} &\frac{1}{\sqrt{3}} &-\frac{1}{\sqrt{2}} \\
-\frac{1}{\sqrt{6}} &\frac{1}{\sqrt{3}} & \frac{1}{\sqrt{2}}
\end{array}
\right)\,.
\end{equation}
%%%%%%%%%%%%%%%%%%%%%%%%%%%%%%%%%%%%%%%%%%%%%%%%%%%%%%%%%%%
%and the
%the neutrino masses result to have the same expressions of \cite{Altarelli:2005yp,Bazzocchi:2007au}. 

%\subsection{The origin of the Cabibbo angle}
%We include now the higher order operators suppressed by powers of the cut-off scale  $\Lambda$. The first terms at order $1/\Lambda^2$
%that change the structure of the  charged fermion mass matrices above are
The same operator that generates the neutrino masses and the lepton mixing matrix,
generates also the Cabibbo angle in the quark sector.
In fact we have that, with the inclusion of the contributions from the $\bf\overline{126}$,
the charged fermion mass matrices of eq.~(\ref{break1}) acquire a very small
diagonal contribution and become%
%%%%%%%%%%%%%%%%%%%%%%%%%%%%%%%%%%%%%%%%%%%%%%%%%%%%%%%%%%%%%%%%%%%%%%%%%%%%
\footnote{We notice that the ratio of the vev of the doublets in the $\bf\overline{126}$
can have a different value from the one in the ${\bf 10}$, for this reason we put the
index $f$ to the $\rho$ parameter, with the relation $\rho^D=\rho^E$.}
%%%%%%%%%%%%%%%%%%%%%%%%%%%%%%%%%%%%%%%%%%%%%%%%%%%%%%%%%%%%%%%%%%%%%%%%%%%%
$$
M^f=\,\frac{m_3^f}{3}\left(
\begin{array}{ccc}
1+\rho^f   &1+\delta^f_1      &1+\delta^f_2\\
1+\delta^f_2&1+\rho^f \omega^2 &1+\delta^f_1\\
1+\delta^f_1&1+\delta^f_2      &1+\rho^f \omega
\end{array}
\right)\,.
$$
%%%%%%%%%%%%%%%%%%%%%%%%%%%%%%%%%%%%%%%%%%%%%%%%%%%%%%%%%%%%%%%%%%%%%%%%%%
In the basis rotated by $\tilde{U}_\omega$ of eq. (\ref{eq:U}),
namely $\tilde{M}^{f}\equiv \tilde{U}_\omega^\dagger\,M^f_{eff}\, \tilde{U}_\omega$, 
the charged fermion  mass matrices are now given by
%%%%%%%%%%%%%%%%%%%%%%%%%%%%%%%%%%%%%%%%%%%%%%%%%%%%%%%%%%%%
\begin{eqnarray}
 \tilde{M}^{f}&=& 
{\,\left(
\begin{array}{ccc}
m^f_1           & \tilde\rho^f \omega^2 & 0 \\
 0              &  m^f_2          & \tilde\rho^f \omega^2 \\
\tilde\rho^f \omega^2 &    0            &  m^f_3 
\end{array}
\right)}
\end{eqnarray}
%%%%%%%%%%%%%%%%%%%%%%%%%%%%%%%%%%%%%%%%%%%%%%%%%%%%%%%%%%%
where $\tilde\rho^f=m_3^f/3 \rho^f$.
Let's assume that the $\tilde\rho^f$ are small arbitrary parameters of 
order $m_3^f O(\lambda^5)$ , where 
$\lambda$ is the Cabibbo angle. The crucial point is that this 
assumption has the consequences that our operators %the higher order operators
give negligible effects in the down and charged lepton sectors, since 
for the down and charged leptons we have
$(m_1^{D,L},m_2^{D,L},m_3^{D,L})\sim(\lambda^4,\lambda^2,1)$ and 
$\tilde{M}^{D,L}$ may be considered diagonals. On the contrary for the 
up quarks we have that 
$(m_1^{U},m_2^{U},m_3^{U})\sim(\lambda^7,\lambda^4,1)$ 
and therefore the off-diagonal entry (1,2) cannot
be neglected: the matrix $\tilde{M}^{U}$ is diagonalized by a
rotation in the 12 plane with $\sin\theta_{12}\approx \lambda$.
This rotation produces the Cabibbo angle in the CKM.
In fact while $M^D$ is still diagonalized by $U_\omega$, 
we have that $M^U$ is diagonalized by
$V_L^{U \dag}\, U_\omega^\dagger M^U U_\omega \,V^U_R$ where $V_{LR}^U$ 
are unitary matrix, approximatively rotations in the 12 plane, and therefore the CKM 
mixing matrix is  given by
%%%%%%%%%%%%%%%%%%%%%%%%%%%%%%%%%%%%%
$$
V_{CKM}= (V_L^U)^\dagger\, U_\omega^\dagger\,U_\omega\equiv (V_L^U)^\dagger\,.
$$
%%%%%%%%%%%%%%%%%%%%%%%%%%%%%%%%%%%%%%%%%%
The charm and top quark masses are almost unaffected by the corrections
and still are given by $m_2^U$ and $m_3^U$ respectively.
The up quark mass is obtained by tuning the $\tilde\rho^U$.

\section{Conclusions}\label{sec:conclusions}
In this work we addressed the two aspects of the flavor puzzle:
the charged fermion mass and the mixing hierarchies.
Following the idea that the mass hierarchy and large mixing angles are
not originated at the same step in the symmetry breaking pattern,
we introduced a GUTF $SO(10)\times(SU(3) \times U(1))^F$ model.

On one hand, a democratic structure for the charged fermion mass matrices
arises from the vev of a scalar that transforms as a ${\bf \bar{6}}$
under the flavor group $SU(3)^F$.
In this way the hierarchy between the 3rd family charged fermion
masses and the others two is explained in a natural way.
When the flavor group is dynamically broken, 
the CKM is given by an undetermined rotation in the $1-2$,
while neutrino are massless and the lepton mixing is undetermined.

On the other hand, the explicitly breaking of $SU(3)^F$ into $A_4$
generates automatically the first and second family charged fermion masses
$m_{1,2}\ll m_3$. However, in order to fit the hierarchy
between the masses of the first and second families, we require a tuning.

Finally, the same operators generate the neutrino masses, the large mixing
lepton angle and the Cabibbo angle.
In fact, assuming that the light neutrino Yukawa interactions come 
from the couplings with an $A_4$ singlet and an $A_4$ triplet
that acquires vev in the direction $(0,0,1)$, we have showed that
the lepton mixing matrix is the tribimaximal one.
In particular in our model these operators give corrections to the
entry $(1,2)$ of all charged fermion mass matrices.
If the ratio between this correction and $m_c$ is of the order
of the Cabibbo angle $\lambda$, we obtain that a rotation
of order $\lambda$ in the 12 plane appears in the up mass matrix.
However the down and charged lepton mass matrices are almost unaffected 
by such corrections. This mismatching gives up the Cabibbo angle in the
quark sector as a net result.

\acknowledgments

One of us (M.P.) acknowledge the kind hospitality of the Murcia University and the
IFIC in Valencia for kind hospitality. 
M.P., E.T-L, and S.M. thank the Milan University for kind hospitality during the
final part of the work. 
We acknowledge a MEC-INFN grant, Fundacion Seneca (Comunidad Autonoma de Murcia)
grant, a CYCIT-Ministerio of Educacion (Spain) grant.
Work partially supported by MEC grants FPA2005-01269 and FPA2005-25348-E, 
by Generalitat Valenciana ACOMP06/154.
This work has been supported by the Spanish Consolider-Ingenio 2010
Programme CPAN (CSD2007-00042).

\bibliographystyle{h-elsevier2}
%\bibliographystyle{prbbib}
%\bibliography{refmt2}

\newpage
%%%%%%%%%%%%%%%%%%%%%%%%%%%%%%%%%%%%%%%%%%%%%%%%%%%%%%%%%%%%%%%%%%%%%%
\begin{table}[t]
\begin{center}
\begin{tabular}{|c||c|c|c|c|c|c|c|c||c|c|c|c|c||c||c|c|c|c|c|}
\hline
%\multicolumn{1}{|c}{}
%&\multicolumn{8}{||c|}{$MSSM$ fields}
%&\multicolumn{1}{c}{}
%&\multicolumn{9}{|c|}{Fields of the explicit breaking $SU(3)^F \times U(1)^F\to A_4$ }\\
%\hline
\multicolumn{1}{|c||}{}
&$L$  
&$Q$
&$N$
&$E$
&$U$
&$D$
&$H_u$
&$H_d$
&$\overline{126}$
&$45_A$
&$45_B$
&$45_C$
&$45_D$
&$\Phi$   
&$\phi$
&$\tilde\phi$
&$\zeta_S$
&$\zeta_T$\\
\quad&\quad\quad&\quad\quad&\quad\quad&\quad\quad&\quad\quad&\quad\quad
&\quad\quad\quad&\quad\quad\quad&\quad\quad\quad\quad&\quad\quad\quad\quad
&\quad\quad\quad\quad&\quad\quad\quad\quad&\quad\quad\quad\quad
&\quad\quad\quad\quad&\quad\quad\quad\quad&\quad\quad\quad\quad
&\quad\quad\quad\quad&\quad\quad\quad\quad\\
\hline
$SO(10)$&
\multicolumn{6}{|c|}{{\bf 16}}
&\multicolumn{2}{|c||}{{\bf 10}}
&${\bf \overline{126}}$&{\bf 45}&{\bf 45}&{\bf 45}&{\bf 45}
&{\bf 1}
&{\bf 1}&{\bf 1}&{\bf 1}&{\bf 1}\\
%\\
\hline
$SU(3)^F$&
\multicolumn{6}{|c|}{${\bf 3}$}&
\multicolumn{2}{|c||}{${\bf 1}$}
&${\bf 1}$
&${\bf 1}$
&${\bf 1}$
&${\bf 1}$
&${\bf 1}$
&{$\overline{\bf 6}$}
&\multicolumn{4}{|c|}{ }\\
$U(1)^F$&
\multicolumn{6}{|c|}{$-\frac{1}{2}\left(\delta +\chi +\omega\right)$}&
\multicolumn{2}{|c||}{$\omega-\rho-\sigma$}
&$\omega+\chi$%$-\alpha$
&$\sigma-\beta$%$-\alpha\alpha$
&$\beta + \chi$
&$\chi$
&$\delta$
&$\delta+\rho$%$+\alpha\alpha$
&$\rho-\chi$%$+\alpha\alpha$
&$\rho+\sigma$
&$\delta$%$+\alpha$
&$\delta$%$+\alpha$
\\
$A_4$&
\multicolumn{6}{|c|}{{\bf 3}}
&\multicolumn{2}{|c||}{{\bf 1}}
&${\bf 1}$
&${\bf 1}$
&${\bf 1}$
&${\bf 1}$
&${\bf 1}$
&
&${\bf 3}$
&${\bf 3}$
&${\bf 1'}$
&${\bf 3}$\\
\hline
\end{tabular}
\caption{The field content of the model. With this charge assignment, all the allowed operators
are the only ones in our Lagrangian of eqs. (\ref{eq:LagU3}-\ref{eq:LagA4}), as explained in the text.
%For simplicity we will put $\alpha=0=\alpha\alpha$.
}
\label{tab:tab1}
\end{center}
\end{table}
%%%%%%%%%%%%%%%%%%%%%%%%%%%%%%%%%%%%%%%%%%%%%%%%%%%%%%%%%%%%%%%%%%
\begin{table}[tbh]
\begin{center}
\begin{tabular}{|c|c|c|c|c|}
\hline
\quad\ \quad\ & \quad $X$ \quad\ & \quad $Y$ \quad\ & $B-L$ &\ \ $T_{3R}$\ \ \\
\hline
$Q$    & 1 & 1/3& 1&0   \\
$U^c$  & 1 &-4/3&-1&1/2 \\
$D^c$  & -3& 2/3&-1&-1/2\\
$L$    & -3&  -1&-3&   0\\
$E^c$  & 1 &   2& 3&-1/2\\
$N^c$& 5 &   0& 3&1/2 \\
\hline
\end{tabular}
\caption{$U(1)$ gauge quantum numbers for the low energy matter fields \cite{Anderson:1993fe}.}\label{tab:tab2}
\end{center}
\end{table}
%%%%%%%%%%%%%%%%%%%%%%%%%%%%%%%%%%%%%%%%%%%%%%%%%%%%%%%%%%%%%%%%%%%%
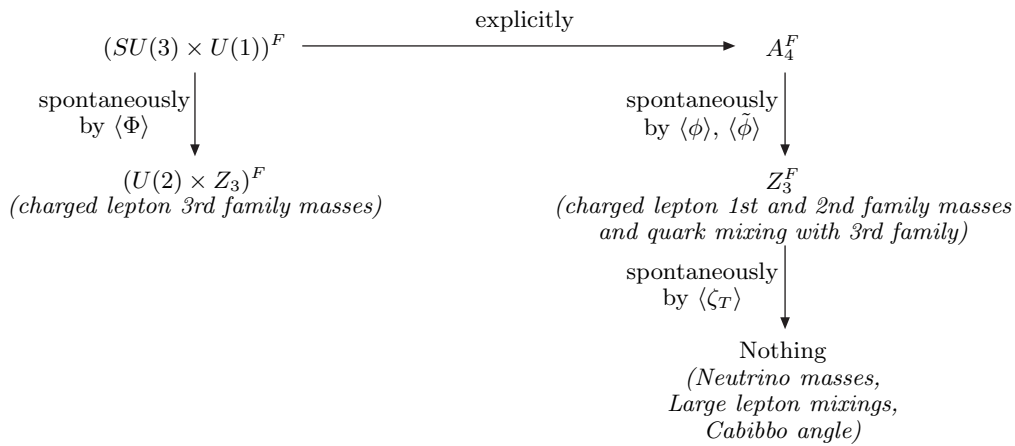
\begin{figure}[bht]
\begin{center}
\begin{picture}(260,200)(0,10)
\Text(0,180)[c]{$(SU(3)\times U(1))^F$}
\LongArrow(40,180)(200,180)
\Text(105,190)[l]{explicitly}
\Text(220,180)[c]{$A_4^F$}
\LongArrow(0,170)(0,140)
\Text(-30,160)[c]{spontaneously}
\Text(-30,150)[c]{by $\langle \Phi \rangle$}
\Text(0,130)[c]{$(U(2)\times Z_3)^F$}
\Text(0,120)[c]{{\em (charged lepton 3rd family masses)}}
\LongArrow(220,170)(220,140)
\Text(190,160)[c]{spontaneously}
\Text(190,150)[c]{by $\langle \phi \rangle$, $\langle \tilde\phi \rangle$}
\Text(220,130)[5]{$Z_3^F$}
\Text(220,120)[c]{{\em (charged lepton 1st and 2nd family masses}}
\Text(220,110)[c]{{\em and quark mixing with 3rd family)}}
\LongArrow(220,105)(220,75)
\Text(190,95)[c]{spontaneously}
\Text(190,85)[c]{by $\langle \zeta_T \rangle$}
\Text(220,65)[5]{Nothing}
\Text(220,55)[c]{{\em (Neutrino masses,}}
\Text(220,45)[c]{{\em Large lepton mixings,}}
\Text(220,35)[c]{{\em Cabibbo angle)}}
%
%\LongArrow(220,45)(220,25)
%\Text(190,35)[c]{spontaneously}
%\Text(220,15)[c]{Nothing}
%\Text(220,5)[c]{{\em (Sub-dominant quark mixing angles)}}
%
\end{picture}
\caption{Diagrammatic representation of the flavor symmetry structure of the model.
The upper horizontal arrow indicates the explicit global symmetry breaking
$SU(3)^F\to A_4^F$ due to the Yukawa terms induced by a hidden sector.
The other arrows show the spontaneous breaking.
The hierarchy among the masses is not directly related to the mixing 
angles.}
\end{center}
\end{figure}
%%%%%%%%%%%%%%%%%%%%%%%%%%%%%%%%%%%%%%%%%%%%%%%%%%%%%%%%%%%%%%%%%%%%%%%%%%%%
\def\figuraNEW{
\begin{figure}[htb]
\begin{center}
\begin{picture}(260,210)(0,40)
\Text(0,230)[c]{$(SU(3) \times U(1))^F\times SO(10)$}
\LongArrow(40,230)(190,230)
\Text(105,240)[l]{explicitly}
\Text(220,230)[c]{$A_4^F\times SO(10)$}
\LongArrow(0,220)(0,190)
\Text(40,210)[c]{spontaneously}
\Text(40,200)[c]{by $\langle {\bf 10} \rangle$}
\Text(0,180)[c]{$(SU(3) \times U(1))^F\times G_{SM}$}
%\Text(0,170)[c]{{\em (charged fermion 3rd family masses)}}
%
\LongArrow(220,220)(220,190)
\Text(180,210)[c]{spontaneously}
\Text(180,200)[c]{by $\langle {\bf 10} \rangle$}
\Text(220,180)[5]{$A_4^F\times G_{SM}$}
%\Text(220,170)[c]{{\em (charged fermion 1st and 2nd family masses and 
%mixing)}}
%
\LongArrow(0,170)(0,140)
\Text(-50,160)[c]{spontaneously}
\Text(-50,150)[c]{by $\langle \Phi \rangle$}
\Text(0,130)[c]{$(U(2)\times Z_3)^F\times G_{SM}$}
\Text(0,120)[c]{{\em (charged fermion 3rd family masses)}}
\LongArrow(220,170)(220,140)
\Text(270,160)[c]{spontaneously}
\Text(270,150)[c]{by $\langle \phi \rangle$}
\Text(220,130)[5]{$Z_3\times G_{SM}$}
\Text(220,120)[c]{{\em (charged lepton 1st and 2nd family masses and mixing)}}
\Line(0,110)(0,65)
\LongArrow(0,65)(190,65)
\Text(-50,95)[c]{spontaneously by}
\Text(-50,85)[c]{higher order operators}
\LongArrow(220,115)(220,75)
\Text(270,105)[c]{spontaneously}
\Text(270,95)[c]{by $\langle \phi' \rangle$ and}
\Text(270,85)[c]{higher order operators}
\Text(220,65)[5]{$G_{SM}$}
\Text(220,55)[c]{{\em (Neutrino masses and mixing, Cabibbo angle)}}
\Text(220,45)[c]{{\em (Sub-dominant quark mixing angles)}}
\end{picture}
\end{center}
\caption{Diagrammatic representation of the flavor and gauge symmetry
structure of the model.
The upper horizontal arrow indicates the explicit global symmetry breaking
$SU(3)^F\to A_4$ due to the Yukawa terms induced by
a hidden scalar sector. The other arrows show the spontaneous
breaking.
The hierarchy among the masses is not directly related to the mixing 
angles.}
\label{fig:symmetries2}
\end{figure}
}
%%%%%%%%%%%%%%%%%%%%%%%%%%%%%%%%%%%%%%%%%%%%%%%%%%%%%%%%%%%%%%%%%%%%%%

\end{document}